\documentclass[aps,prd,twocolumn,groupedaddress,showpacs,showkeys]{revtex4-1}
\usepackage{bm}% bold math
\usepackage{amssymb,amsmath,latexsym,amsfonts}
\begin{document}

%\preprint{APS/123-QED}
\title{Speed of sound in a Bose-Einstein condensate}

\author{A. Camacho}
 \email{acq@xanum.uam.mx} \affiliation{Departamento de F\'{\i}sica,
 Universidad Aut\'onoma Metropolitana--Iztapalapa\\
 Apartado Postal 55--534, C.P. 09340, M\'exico, D.F., M\'exico.}

\begin{abstract}
In the present work we determine the speed of sound in a
Bose--Einstein condensate confined by an isotropic harmonic
oscillator trap. The deduction of this physical parameter is done
resorting to the $N$--body Hamiltonian operator. The
single--particle eigenfunctions that have been employed in this
formalism are those stemming from the corresponding harmonic
oscillator potential, and an expression for the dependence of this
speed on the temperature is also deduced. These functions are used
in the calculation of the scaterring length, etc. The situation for
a Bose--Einstein condensate of sodium is evaluated and the
corresponding speed of sound is obtained and compared against the
known experimental outcomes. The possibility that the solution, to
the existing discrepancy between experiment and theoretical
predictions, could be given by the Zaremba--Nikuni--Griffin
formalism is also explored.
 \end{abstract}

\maketitle

%****************************************************************************

\section{Introduction}

A fundamental aspect in the context of quantum gases is related to
the analysis of their collective excitations. Collective modes in
uniform quantum fluids can be categorized in two different realms,
namely, collisionless and hydrodynamic modes. The former are
associated to the dynamic self--consistent mean fields, whereas, the
latter emerge as a consequence of the properties of the interactions
\cite{Griffin1}. The study of collective excitations could provide a
deep insight into some physical characteristics of these systems. In
other words, the study of the excitations in a Bose--Einstein
condensate (BEC) opens up a new realm in the comprehension of the
structure of quantum fluids, at least in the regime of dilute
quantum fluids. Clearly, in this direction there are several
properties to be considered; among them we may find the role that
trapping potentials and interactions among the constituents of the
system play in the determination of certain features of a BEC. At
this point let us focus our attention on the consequences of binary
interactions upon some features of a BEC. In some cases, for
instance, $^{87}Rb$ or $^{23}Na$, the size of the corresponding
condensate suffers an enlargement due to the presence of the
two--body repulsive forces. A further effect of these repulsive
interactions is the fact that the central density of a non--ideal
BEC at very low temperatures can be two or three orders of magnitude
higher than the one associated to an ideal BEC \cite{Dalfovo}, i.e.,
when the interaction among the particles of the gas is switched off.

A knowledge of the speed of sound opens up several possibilities.
For instance, the study of correlated momentum excitations in the
many--body condensate wave function \cite{Stamper, Papp}. As a
further comment let us mention that the dispersion relation for
elementary excitations has a very different form when the
possibility of an interaction among the particles is considered or
not \cite{Stringari1}. Since binary collisions are not a frequent
event in a BEC it is a little bit surprising that the concept of
interaction plays a primordial role in the determination of some
physical features. The answer to this interrogant stems from the
large coherent mean field associated to a BEC \cite{ZNG}. These
comments lead us to conclude that the comprehension of the speed of
sound in a condensate defines an issue that bears an important
physical relevance. In this context let us mention that it is
believed that Gross--Pitaevskii mean field formalism captures the
most essential properties associated to the ground state of a BEC
\cite{Edwards}. Of course, the comparison of the theoretical
predictions against the experimental outcomes provides a test of the
validity of the mean field formalism, at least indirectly.

Usually the deduction of the speed of sound in a BEC is done
resorting to the linearized time--independent Gross--Pitaevskii
equation in the so--called Tomas--Fermi limit. Afterwards, this
expression is cast in the form of two quantum hydrodynamic equations
(one for the density fluctuations and the second one for the
velocity). Finally, the velocity is eliminated from the
aforementioned equations and a differential equation for the density
fluctuations is obtained \cite{Zaremba1}. Clearly, the
aforementioned procedure does not exhaust all the possible manners
in which the speed of sound can be deduced. Another way starts from
the N--body Hamiltonian in the second quantization formalism,
introduces the Bogoliubov approximation and the resulting
Hamiltonian is diagonalized by means of the Bogoliubov canonical
transformation. The energy of the ground state of this diagonalized
Hamiltonian allows us to calculate the pressure and speed of sound
of the corresponding BEC \cite{Ueda}. A crucial point in all these
approaches is the Mean Field Theory (MFT) \cite{Chaikin}. This
formalism requires the use of several assumptions, one of them is
related to the introduction of single--particle wave functions, a fact
that can be tracked down to the Gibbs--Bogoliubov--Feynman equation
\cite{GBF, Prato}. The choice of these aforementioned
single--particle eigenfunctions is a consequence of a minimization
procedure, nevertheless, in some cases free--particles
eigenfunctions \cite{Ueda}, in other situations the eigenfunctions
related to the particular trapping potential are employed
\cite{Stringari1}, without proving if they are the minimizing case.

In the present work we determine the speed of sound in a
Bose--Einstein condensate confined by an isotropic harmonic
oscillator trap. The deduction of this physical parameter is done
resorting to the $N$--body Hamiltonian operator. The
single--particle eigenfunctions that have been employed in this
formalism are those stemming from the corresponding harmonic
oscillator potential, and an expression for the dependence of this
speed on the temperature is also deduced. These functions are used
in the calculation of the scattering length, etc. The situation for
a Bose--Einstein condensate of sodium is evaluated and the
corresponding speed of sound is obtained and compared against the
known experimental outcomes. Finally, we must mention that there is
a discrepancy, the one emerges from the comparison between some of
the extant theoretical predictions against the experimental
outcomes. Here we discuss also the possibility that the solution to
this discrepancy could be given by the Zaremba--Nikuni--Griffing
formalism.

\section{Mean Field Theory and Speed of Sound in a BEC}

Our starting point is the $N$--body Hamiltonian operator, in
addition we assume that the collisions in the gas are, mainly,
two--body interactions, this result is a consequence of the fact
that we introduced, as additional condition, a dilute gas
\cite{Ueda}. Under these restrictions the aforementioned Hamiltonian
reads

\begin{eqnarray} \hat{H} =-\frac{\hbar^2}{2m}\sum_{\alpha,\beta}<\alpha\vert\nabla^2\vert\beta>
\hat{a}_{\alpha}^{\dagger}\hat{a}_{\beta}\nonumber\\
\frac{1}{2}\sum_{\alpha,\beta, \gamma,\epsilon}<\alpha,\beta\vert\
V\vert\gamma,\epsilon>
\hat{a}_{\alpha}^{\dagger}\hat{a}_{\beta}^{\dagger}\hat{a}_{\gamma}\hat{a}_{\epsilon}\nonumber\\
+\sum_{\alpha}<\alpha\vert\ V_{(e)}(\vec{r})\vert\beta>
\hat{a}_{\alpha}^{\dagger}\hat{a}_{\beta}. \label{Ham1}
\end{eqnarray}

In this last expression we have the following terms

\begin{eqnarray} <\alpha\vert\nabla^2\vert\beta> = \int
u^{\star}_{\alpha}(\vec{r})\nabla^2u_{\beta}(\vec{r})d^3r,
\label{Ham2}
\end{eqnarray}

\begin{eqnarray} <\alpha,\beta\vert\
V\vert\gamma,\epsilon> = \int
u^{\star}_{\alpha}(\vec{r})u^{\star}_{\beta}(\vec{r})V(\vec{r})u_{\gamma}(\vec{r})u_{\epsilon}(\vec{r})d^3r,
\label{Ham3}
\end{eqnarray}

\begin{eqnarray}
<\alpha\vert\ V_{(e)}\vert\beta>= \int
u^{\star}_{\alpha}(\vec{r})V_{(e)}(\vec{r})u_{\beta}(\vec{r})d^3r.
\label{Ham4}
\end{eqnarray}

Here $V(r)$ denotes the two--body potential, whereas
$V_{(e)}(\vec{r})$ depicts the trapping potential. Clearly, we have,
explicitly, assumed that only two--body interactions are relevant
for our case. An explanation for this approximation is related to
the fact that the systems employed in condensation are always,
sufficiently, dilute. Let us explain this fact; usually, particle
separations, in the case of alkali atom vapours, have an order of
magnitude of, approximately, $10^2nm$, whereas, the scattering
length (here denoted by $a$) is two orders of magnitude smaller,
$a\approx 100a_0$, being $a_0$ the Bohr radius \cite{Pethick}. These
comments entail that an $(n+1)$--body collision is less probable
than an $n$--body collision, i.e., we may keep only two--body
collisions.

At this point no restriction upon the two--body interaction has been
imposed. The next step, in this direction, concerns the introduction
of some of the postulates of MFT. Indeed, the assumption of very low
temperature implies that the $s$--wave approximation can be used and
the interatomic potential can be described by the so--called
pseudo--potential. At low temperatures one of the features of
two--body interactions is the emergence of the concept of scattering
length as a fundamental idea \cite{Pethick}. This last comment does
not allow us to evaluate (\ref{Ham3}). Indeed, we must know the set
of single--particle functions, i.e., $\{u_{\beta}(\vec{r})\}$. It
has to be stressed that the introduction of single--particle
wavefunctions in (\ref{Ham1}) already implies the introduction, at
least partially, of MFT. This last assertion can be understood
noting that MFT assumes that the $n$--body symmetrized wavefunction
can be replaced by a $1$--body case where the bridge between these
two cases is given by the deduction (via a minimization process
stemming from the Gibbs--Bogoliubov--Feynmann equation) of a set of
single--particle wavefunctions which lead to the definition of the
so--called Mean Field Hamiltonian \cite{GBF}.

In the context of the calculations about the speed of sound in BEC
usually two choices are made, namely, (i) The eigenfunctions of a
three-dimensional harmonic oscillator \cite{Zaremba1}; (ii) Free
particle wavefunctions \cite{Ueda}. These choices are made without
checking if they correspond to the minimum required by the
Gibbs--Bogoliubov--Feynmann formalism. Of course, the first choice,
for the case of a BEC trapped by a three--dimensional harmonic
oscillator, seems to be a good conjecture since it reflects the
symmetry of the trap.

\section{Speed of Sound and Non--vanishing Temperature}

In the present work our trap will be an isotropic three-dimensional
harmonic oscillator (with a frequency equal to $\omega$) and the
corresponding eigenfunctions will be our choice for single--particle
wavefunctions to be introduced in (\ref{Ham1}). Since we consider
the limit of very low temperatures we assume that only the first
excited state is populated. In other words, ground state and first
excited state are the only states populated. For an ideal BEC (no
interactions among the particles of the gas) the number of particles
in excited states (for temperatures below the condensation
temperature) is a function of the trapping potential, namely, $N_e =
N \bigl[1- \Bigl(\frac{T}{T_c}\Bigr)^{\alpha}\Bigr]$, here the
potential is represented by $\alpha$ \cite{Pethick}. If we consider
the presence of repulsive two--body interactions, then the number of
particles in the ground and first excited state, $N_0$ and $N_e$,
respectively, for temperatures below the condensation temperature
($T_c$), are given by

\begin{eqnarray}
 N_0 = N \Bigl[1- \Bigl(\frac{T}{T_c}\Bigr)^3 + \frac{8}{3}\sqrt{\frac{a^3N}{V\pi}}\Bigr]. \label{Exce1}
\end{eqnarray}

\begin{eqnarray}
 N_e = N \Bigl[\Bigl(\frac{T}{T_c}\Bigr)^3 -\frac{8}{3}\sqrt{\frac{a^3N}{V\pi}}\Bigr]. \label{Exce133}
\end{eqnarray}

In this last expression $a$ denotes the scattering length, and $V$
the volume of the BEC. In the present case, in which there is no
container of volume $V$, but a trap confines the gas, the definition
of $V$ requires a sound explanation. In this context we recall that
a one--dimensional harmonic oscillator, whose frequency reads
$\omega$, acting upon a quantum particle of mass $m$ defines a
length parameter

\begin{eqnarray}
 l = \sqrt{\frac{\hbar}{m\omega}}. \label{Exce2}
\end{eqnarray}

The ground state wavefunction reads

\begin{eqnarray}
\psi(x) = \sqrt{\frac{1}{l\sqrt{\pi}}}\exp\{-\frac{x^2}{2l^2}\}.
\label{Exce3}
\end{eqnarray}

The size of this one--dimensional system will be defined by the
value $x= x_l$ such that $\psi(x_l) = \psi(x=0)e^{-1} $, i.e., $x_l
= \sqrt{2}l$.

Under these conditions a three--dimensional system has a volume
given by (\ref{Ham1})

\begin{eqnarray}
V = 2^{3/2}l^3. \label{Exce4}
\end{eqnarray}

With these arguments we now proceed to calculate (\ref{Ham1}).

The first term to be addressed involves the kinetic energy. In order
to do this we cast the operator $\hat{p}^2$ as a function of the
creation and annihilation operators \cite{Sakurai}

\begin{eqnarray}
\frac{\hat{p}^2}{2m}=
\frac{\hbar\omega}{4}\Bigl[\hat{a}^{\dagger}\hat{a}+
\hat{a}\hat{a}^{\dagger} -(\hat{a}^{\dagger})^2 - (\hat{a})^2\Bigr].
\label{Expl1}
\end{eqnarray}

The corresponding term $<\alpha\vert\nabla^2\vert\beta>$ has to be
calculated over all the possibilities for $\alpha$ and $\beta$.
Since, by hypothesis, we have assumed that only the ground and first
excited states are populated, then $<\alpha\vert\nabla^2\vert\beta>$
does not vanish if and only if the following two conditions are
fulfilled: (i) $\alpha$ or $\beta$ represent the ground or first
excited states; (ii) $\alpha = \beta$. This is no new restriction.
Indeed, according to the rules that the annihilation and creation
operators satisfy notice that
$<\alpha\vert\hat{a}^{\dagger}\hat{a}\vert\beta> =
\gamma\delta_{(\alpha+1,\beta-1)}$, where $\gamma$ is a real number.
It is readily seen that this last expression does not vanish if the
following two conditions are fulfilled: (i) $\gamma\not=0$, and (ii)
$\alpha -1= \beta+1\Rightarrow\alpha = \beta+2$. This last
conclusion entails, since only the ground and first excited states
are populated, that $<\alpha\vert\hat{a}^{\dagger}\hat{a}\vert\beta>
= 0$ if $\alpha\not=\beta$. These arguments allow us to deduce, very
easily, that if $\vert\psi_N>$ denotes the wave function of our
$N$--body system then

\begin{eqnarray}
<\psi_N\vert\frac{\hat{p}^2}{2m}\vert \psi_N>=
\frac{3\hbar\omega}{4}N_0 + \frac{5\hbar\omega}{4}N_e. \label{Expl2}
\end{eqnarray}

Resorting to (\ref{Exce1}) we may cast this last expression in the
following form

\begin{eqnarray}
<\psi_N\vert\frac{\hat{p}^2}{2m}\vert \psi_N>=
\frac{3\hbar\omega}{4}N \Bigl[1- \Bigl(\frac{T}{T_c}\Bigr)^3 +
\frac{8}{3}\sqrt{\frac{a^3N}{V\pi}}\Bigr] \nonumber\\
+\frac{5\hbar\omega}{4}N\Bigl[\Bigl(\frac{T}{T_c}\Bigr)^3 -
\frac{8}{3}\sqrt{\frac{a^3N}{V\pi}}\Bigr]. \label{Expl3}
\end{eqnarray}

This last result provides us with an expression that shows a
temperature dependence, for the case in which $T\leq T_c$. We now
consider (\ref{Ham3}). In the calculation of this term we will
assume that the involved particles are in the ground state. Clearly,
energy has to be conserved. This condition entails that if we denote
by a superindex the values of the occupation numbers before and
after the collision, $b$ and $a$, respectively, then energy
conservation entails for our two involved particles

\begin{eqnarray}
\sum_{i=1}^2\Bigl[^{(i)}n_x^{(a)} + ^{(i)}n_y^{(a)} +
^{(i)}n_z^{(a)}\Bigr] = \nonumber\\
\sum_{i=1}^2\Bigl[^{(i)}n_x^{(b)} + ^{(i)}n_y^{(b)} +
^{(i)}n_z^{(b)}\Bigr].\label{Cal1}
\end{eqnarray}

In the case in which the eigenfunctions of single--particle are the
free--particle functions, the evaluation of the corresponding
integral is done resorting to momentum conservation \cite{Ueda}. In
this direction we find our sought term

\begin{eqnarray}
<0,0\vert V\vert 0,0> = \frac{32}{\sqrt{\pi}l^3}\int_0^{\infty}
r^2V(r)\exp\{-\frac{r^2}{l^2}\}dr.\label{Ham13}
\end{eqnarray}

Here $r$ denotes the relative distance between the involved
particles. This last expression leads us to the definition of the
concept of scattering length

\begin{eqnarray}
a = \frac{m}{4\pi\hbar^2}\int_0^{\infty}
r^2V(r)\exp\{-\frac{r^2}{l^2}\}dr.\label{Ham23}
\end{eqnarray}

The usual situation is proportional to the integral of the
interaction potential \cite{Pethick}. We now deduce the energy of
this system, as a function of the temperature, in the usual manner
\cite{Ueda}

\begin{eqnarray}
E(T) = \frac{3\hbar^2}{2mV^{2/3}}N\Bigl[1 -
\Bigl(\frac{T}{T_c}\Bigr)^3 + \frac{8}{3}\sqrt{\frac{a^3N}{V\pi}}
\Bigr] +
\nonumber\\
\frac{5\hbar^2}{2mV^{2/3}}N\Bigl[\Bigl(\frac{T}{T_c}\Bigr)^3
-\frac{8}{3}\sqrt{\frac{a^3N}{V\pi}}\Bigr]+ \nonumber\\
\frac{2\pi
a\hbar^2}{mV}N^2\Bigl[3 -
\frac{4}{3}\sqrt{\frac{a^3N}{V\pi}}\Bigr].\label{Ham43}
\end{eqnarray}

The pressure ($P(T)$) reads

\begin{eqnarray}
P(T) = \frac{\hbar^2}{mV^{5/3}}N +
\frac{2\hbar^2}{3mV^{5/3}}N\Bigl[\Bigl(\frac{T}{T_c}\Bigr)^3
\nonumber\\
-\frac{17}{3}\sqrt{\frac{a^3N}{V\pi}} + \frac{2\pi
a\hbar^2}{mV^2}N^2\Bigl[3 -
2\sqrt{\frac{a^3N}{V\pi}}\Bigr].\label{Ham53}
\end{eqnarray}

The speed of sound ($c_s(T)$) is given by

\begin{eqnarray}
c^2_s(T) = \frac{5\hbar^2}{3m^2V^{2/3}} +
\frac{10\hbar^2}{3m^2V^{2/3}}\Bigl[\Bigl(\frac{T}{T_c}\Bigr)^3
\nonumber\\
-\frac{68}{15}\sqrt{\frac{a^3N}{V\pi}} + \frac{4\pi
a\hbar^2}{m^2V}N\Bigl[3 -
\frac{5}{2}\sqrt{\frac{a^3N}{V\pi}}\Bigr].\label{Ham63}
\end{eqnarray}

\section{Discussion and conclusions}

Our last expression allows us to predict the speed of sound, as a
function of the temperature, of course, this happens only in the
regime $T\leq T_c$. Notice that our prediction does not include a
dependence upon the amplitude of the disturbance, a fact that
matches with the experimental output \cite{Edwards, Andrews}.

It has already been recognized that many theoretical studies, in the
realm of BEC, ignore the role that the thermally excited atoms play
in the definition of the characteristics of the gas \cite{ZNG}. The
assumption of vanishing temperature is not correct, as a matter of
fact thermodynamics tells us that in a practical sense the
achievement of this temperature is impossible \cite{Pathria}.
Therefore, the deduction of the speed of sound under the assumption
of $T=0$ is an approximation, the one should be improved. A point
that has to be underlined in the present manuscript concerns this
issue. Indeed, a fleeting glimpse at (\ref{Expl3}) tells us that the
second term on the right hand--side takes into account the
contribution to the kinetic energy of the thermal cloud of the
system as a function of the temperature, i.e., we do not assume
$T=0$.

In the context of the assumptions here accepted, of course, we have,
as mentioned before, discarded the possibility of having, in this
kinetic term, transitions between excited states and the ground
state. In order to have this case we must consider that not only the
first excited state is populated, but also higher states. From this
last comment we expect a very small contribution to the speed of
sound stemming from this neglected possibility. An additional
simplification has been introduced, but now in relation with
(\ref{Ham3}). If we consider (\ref{Ham13}) we, immediately, notice
that we have only considered interactions between particles in the
ground state with particles in the same state. Of course, more
possibilities are present, for example, a particle in the ground
state might interact with a particle in the first excited state.
Since the number of particles in the ground state is much larger
than those in excited states we expect that the probability of
having a ground state--ground state interaction is larger than
having a ground state--first excited state interaction, and this
last one is larger than the first state--first state interaction. In
this sense has to be understood the assumption.

Let us now confront our theoretical prediction against the extant
experimental results. We will consider the case of a BEC comprised
by sodium atoms. This choice is done since we have already
experimental results in this direction; $N=5\times 10^6$,
$n=10^{21}m^{-3}$, $T_c = 2\times 10^{-6}$K, $m=35.2\times
10^{-27}$Kg, $l\sim 10^{-2}m$ \cite{Andrews}. Finally, the
scattering length, for sodium, has already been measured
\cite{Tiesinga}, namely, $a= 2.75\times 10^{-9}$m. We need also an
assumption for our temperature, here we assume $T=0.9T_c$. These
values imply

\begin{eqnarray}
c_s = 2.2\times 10^{-3}m/s.\label{Ham73}
\end{eqnarray}

A careful look at the present measurement outputs entails that our
result is not a bad one \cite{Andrews}, i.e., it provides the
correct order of magnitude. The shortcomings of the present
manuscript, at least in the realm of its compatibility with the
experimental results, are also shared by other approaches
\cite{Andrews, Andrews1}. The most intriguing interrogant in this
sense is related to the fact that in the region of the
thermodynamical space in which the assumption of MFT is strongly
satisfied the theoretical prediction has its worst behavior
\cite{Andrews1}. Several conjectures could be put forward, in order
to solve this puzzle. The inclusion of additional terms, for
instance, in connection with the discussion about the number of
significant terms related to (\ref{Ham3}) would, surely, modify the
result. In this direction an alternative way is related to the
deduction of the speed of sound resorting to the generalized
Gross--Pitaevski equation (usually known as Zaremba--Nikuni--Griffin
equation (ZNG) \cite{ZNG}) the one takes into account a coupling
between the condensate and non--condensate components of the
corresponding system. A physical motivation behind this statement
can be found in the fact that sound can be understood as density
waves. Clearly, changes in the density, which involve changes in the
separation among particles, are determined by the interactions among
the constituents of the gas. This explains in a very simple way why
the idea of interaction plays a relevant role in the determination
of the speed of sound. The ZNG equation \cite{ZNG} includes in the
dynamics of a BEC a coupling between the condensate and the
noncondensate part of the gas. We may rephrase this last conjecture
asserting that one of our assumptions, namely, {\it no interaction
between condensate and noncondensate components} could define a
wrong premise in this context. In other words, the present
approaches render the correct order of magnitude for this speed, but
finer details involved in the correct deduction of it could stem
from the use of the ZNG equation.


\begin{thebibliography}{}
\bibitem{Griffin1}
D. A.~Griffin, Wen--Chin Wu, and S. Stringari, Phys. Rev. Lett.
\textbf{78}: 1838--1841 (1997).

\bibitem{Dalfovo} F. Dalfovo et al, Rev. Mod. Phys. \textbf{71},
463--512 (1999).

\bibitem{Stamper}
D. M.~Stamper et al, Phys. Rev. Lett. \textbf{83}: 2876--2879
(1999).

\bibitem{Papp}
S. B.~Papp et al, Phys. Rev. Lett. \textbf{101}: 135301 (2008).

\bibitem{Stringari1}
S.~Stringari, Phys. Rev. Lett. \textbf{77}: 2360--2363 (1996).

\bibitem{ZNG} A.~Griffin, T.~Nikuni, and E.~Zaremba.  \emph{Bose--Condensed Gases at Finite Temperatures}
(Cambridge University Press 2009).

\bibitem{Edwards}
M.~Edwards et al, Phys. Rev. Lett. \textbf{77}: 1671--1674 (1996).

\bibitem{Zaremba1}
E.~Zaremba, Phys. Rev. \textbf{A57}: 518--521 (1998).

\bibitem{Ueda} M.~Ueda, \emph{Fundamentals and New Frontiers of Bose--Einstein Condensation}
(World ScientificPublishing Co. 2010).

\bibitem{Chaikin} P.~M.~Chaikin and T. C.~Lubensky.  \emph{Principles of condensed matter physics}
(Cambridge University Press 2000).

\bibitem{GBF} R. P.~Feynman and A. R.~Hibbs.  \emph{Quantun Mechanics and Path Integrals}
(Mc--Graw--Hill, New York 1965).

\bibitem{Prato}
D.~Prato and D. E.~Barraco, Rev. Mex. Fis. \textbf{42}: 145--150
(1996).

\bibitem{Pethick} C. J.~Pethick and H.~Smith.  \emph{Bose--Einstein Condensation in Dilute Gases}
(Cambridge University Press 2006).

\bibitem{Sakurai}
J. J.~Sakurai, \emph{Modern Quantum Mechanics}  (Addison--Wesley
Publishing Company, New York, 1994).

\bibitem{Andrews}
M. R.~Andrews et al, Phys. Rev. Lett. \textbf{79}: 553--556 (1997).

\bibitem{Pathria}
R. K.~Pathria, \emph{Statistical Mechanics}  (Butterwoth--Heinemann,
Oxford, 1996).

\bibitem{Tiesinga}
E.~Tiesinga et al, Res. Nat. Inst. Stand. Technol \textbf{101}:
505--520 (1996).

\bibitem{Andrews1}
M. R.~Andrews et al, Phys. Rev. Lett. \textbf{80}: 2967 (1998).


\end{thebibliography}
\end{document}